\documentclass[
  amsmath,amssymb,
  preprint,
 ]{revtex4-1}
 
 \usepackage{graphicx}% Include figure files
 \usepackage{dcolumn}% Align table columns on decimal point
 \usepackage{bm}% bold math
 \usepackage[mathlines]{lineno}% Enable numbering of text and display math
 
 \usepackage{pifont}
 
 \usepackage[utf8]{inputenc}
 \usepackage[T1]{fontenc}
 \usepackage{mathptmx}
 \usepackage{etoolbox}
 
 \usepackage[
 colorlinks=true,
 linkcolor=blue,
 anchorcolor=blue,
 citecolor=blue,
 urlcolor=blue
 ]{hyperref}
 
 \makeatletter
 \def\@email#1#2{%
  \endgroup
  \patchcmd{\titleblock@produce}
   {\frontmatter@RRAPformat}
   {\frontmatter@RRAPformat{\produce@RRAP{*#1\href{mailto:#2}{#2}}}\frontmatter@RRAPformat}
   {}{}
 }%
 \makeatother

\begin{document}
 
 \setlength{\parskip}{0pt}
 
 \title{Continuously and widely tunable frequency-stabilized laser based on an optical frequency comb}
 % Force line breaks with \\

 \author{Ze-Min Shen,$^\mathsf{1,\;2,\;3\;}$Xiao-Long Zhou,$^\mathsf{1,\;2,\;3\;}$Dong-Yu Huang,$^\mathsf{1,\;2,\;3\;}$Yu-Hao Pan,$^\mathsf{1,\;2,\;3\;}$Li Li,$^\mathsf{1,\;2,\;3\;}$Jian Wang,$^\mathsf{1,\;2,\;3,\;a)\;}$Chuan-Feng Li,$^\mathsf{1,\;2,\;3,\;a)\;}$ and Guang-Can Guo}

 \affiliation{CAS Key Laboratory of Quantum Information, University of Science and Technology of China, Hefei 230026, China}
 \affiliation{CAS Center For Excellence in Quantum Information and Quantum Physics, University of Science and Technology of China, Hefei 230026, China}
 \affiliation{Hefei National Laboratory, University of Science and Technology of China, Hefei 230088, China}
 \email[$^\mathsf{a)}$ The authors to whom correspondence may be addressed:]{$\;$jwang28@ustc.edu.cn and cfli@ustc.edu.cn}
 
 \date{\today}% It is always \today, today,
 
 \begin{abstract}
   Continuously and widely tunable lasers actively stabilized on a frequency reference are broadly employed in atomic, molecular and optical (AMO) physics. The frequency-stabilized optical frequency comb (OFC) provides a novel optical frequency reference with a broadband spectrum that meets the requirement of laser frequency stabilization. Therefore, we demonstrate a frequency-stabilized and precisely tunable laser system based on it. In this scheme, the laser frequency locked to the OFC is driven to jump over the ambiguity zones, which blocks the wide tuning of the locked laser, and tuned until the mode hopping happens with the always-activated feedback loop. Meanwhile, we compensate the gap of the frequency jump with a synchronized acoustic optical modulator to ensure the continuity. This scheme is applied to an external cavity diode laser (ECDL) and we achieve tuning at a rate of about 7 GHz/s with some readily available commercial electronics. Furthermore, we tune the frequency-stabilized laser only with the feedback of diode current and its average tuning speed can exceed 100 GHz/s. Due to the resource-efficient configuration and the simplicity of completion, this scheme can be referenced and find wide applications in AMO experiments.
 \end{abstract}
 
 \maketitle
 
 \section{Introduction}
 Widely tunable and frequency-stabilized lasers \cite{X1,R2} are commonly needed in atomic, molecular and optical experiments, such as atomic and molecular manipulation \cite{Z2,Z3,Z4}, precision spectroscopy \cite{R5}, optical clock \cite{R6,R7}, gravitational wave detection \cite{R8} and so on. The laser frequency is often stabilized on an optical cavity with the Pound-Drever-Hall (PDH) technique \cite{R9}. However, the optical cavity must be set finely with the mode-matching optics \cite{Z5} and this method needs the process of phase modulation and demodulation. Meanwhile, it is difficult to tune the laser widely based on the optical cavity. One way is to lock the laser and then change the frequency reference by varying the cavity length \cite{Z6,Z7} or the optical index\cite{Z1}. Another way is to lock the sideband of the laser and then tune the sideband with the modulation of an electro-optic modulator (EOM) \cite{R10}. However, the former reduces the stability of the cavity and the latter has a restricted tuning range limited by the operation range of the EOM. In addition, it is complex to lock and tune multiple lasers based on one optical cavity.
 
 The frequency-stabilized OFC is a novel and accurate frequency reference \cite{R11,M1,M2} without long-term drift or sideband modulation for the laser stabilization \cite{R12,Z9,R13}. Within its broad spectrum, there are hundreds of thousands of phase-coherent and evenly spaced narrow comb lines which are locked to optical references or microwave atomic or molecular transitions. However, it is unsuitable to be used directly for some applications, since there are too many comb modes and the distributed power of each mode is too low \cite{R20}. This problem is solved by employing a tunable laser locked on the OFC. The laser frequency is stabilized by locking the beat signal between the laser and the OFC to a radio-frequency (RF) source and then it can be tuned accordingly by changing the working frequency of RF source. Nevertheless, because of the existence of ambiguity zones near the comb mode and the middle position of two adjacent comb modes in the frequency domain, which confuse the locking module and cause a feedback error, the tuning range of the laser frequency with this method is less than half repetition rate of the OFC. A number of solutions eschewing the ambiguity zone have been demonstrated in the previous reports, such as adjusting the carrier-envelope offset frequency or the repetition rate of the OFC \cite{R14,R15,R16,R17,Z8}. Though these methods are convenient and efficient, we are not allowed to manipulate multiple lasers locked on one OFC individually, since their common frequency reference is changed. Then multiple novel methods of laser tuning without adjusting the parameters of the OFC are proposed. For instance, a holding command is applied to the feedback loop and the control signal is frozen simultaneously as the the laser frequency is tuned near the ambiguity zone. Then adding an independent signal to guide the laser frequency to jump over the ambiguity zone can realize the wide laser tuning \cite{R18,R19}. In addition, it can also be realized by producing two different beat signals and then locking and tuning one chosen beat signal which is not close to the ambiguity zone \cite{R20}. Tuning N lasers continuously and independently at a rate of 200 MHz/s is reported with an acousto-optic modulator (AOM) which is employed before the laser is compared by optical heterodyne with the OFC \cite{R21}. Futhermore, tuning over 3 GHz in about one millisecond is demonstrated by ramping freely and then relock the laser to the OFC at the end of each ramp \cite{R22}.
 
 In this paper, we demonstrate a continuous and wide ECDL tuning system which is stabilized on an OFC and achieve a precisely linear tuning at a rate of 7 GHz/s before the mode hopping happens. Furthermore, we turn off the feedback of the piezo-electric transducer which mechanically limits the speed and then reach a tuning rate more than 100 GHz/s only with feedback of the diode current. In this scheme, we add an external pulse signal directly to the error signal and drive the beat signal to jump over the ambiguity zone without suspending the feedback loop. The always-activated feedback loop reduces the complexity of the control system compared with the previous methods \cite{R18,R19}, in which a brief shutoff to the feedback loop is needed. In addition, we compensate the gap of the frequency jump with a synchronized acoustic optical modulator, ensuring there is none of missing point in the laser tuning process. With some simple and readily available electronics, a number of lasers can be locked on one OFC and tuned independently to each other, since there is no need to change the spectrum of the OFC. Thanks to the resource-efficient configuration and the simplicity of completion, this scheme can find wide applications in AMO applications.
 
 \section{Experimental setup}
 
\begin{figure}[h!]
   \centering
   \includegraphics[width=12cm]{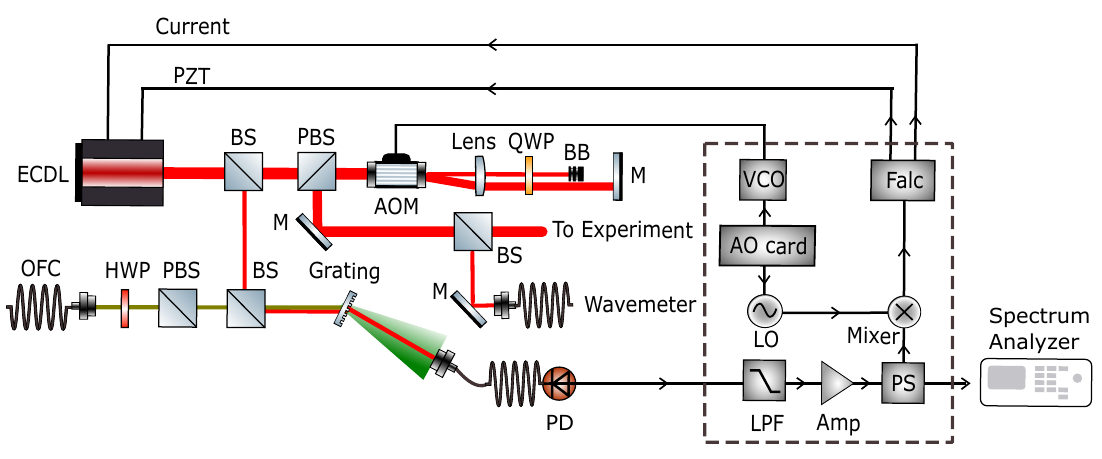}
   \caption{
   \label{setup}
   The experimental setup. The beat signal filtered spatially by the optical grating and single mode fiber is transferred to the electronic setup (dashed box) for stabilizing and tuning the laser frequency. The polarizing beam splitter in the light path of the OFC is used to ensure that its polarization is the same as the diode laser's for the subsequent interference, since the initial output diode laser is horizontally polarized.  
   ECDL: external cavity diode laser; BS: beam splitter; AOM: acousto-optic modulator; OFC: optical frequency comb; HWP: half wavelength plate; QWP: quarter-wave plate; PBS: polarizing beam splitter; BB: beam block; M: mirror; PD: photodetector; LPF: low pass filter; Amp: amplifier; PS: power splitter; LO: local oscillator; AO card: analog output card; VCO: voltage controlled oscillator; Falc: Falc 110, a fast locking module; PZT: piezo-electric transducer. }
 \end{figure}

 As shown in FIG.~\ref{setup}, the experimental setup consists of two parts: the optical setup for detecting the beat signal between the OFC and the ECDL, and the electronic setup for stabilizing and tuning the laser frequency which is shown in the dashed box.
 
 In the optical setup, the frequency of ECDL (DL pro, Toptica) at 780 nm can be controlled by adjusting the laser diode current and the offset voltage of the piezo-electric transducer(PZT). We use a wavemeter (WS7-30 IR, High Finesse and Angstrom) to observe the laser tuning with a measurement resolution of 5 MHz and an absolute accuracy of 30 MHz. The frequency reference, self-referenced and frequency-stabilized OFC (FC1500-250-ULN, Menlo Systems), has a spectrum around 1550 nm and part of it is converted to the second harmonic at 780 nm with a spectral width more than 20 nm. Its repetition rate $f_{rep}$ and carrier-envelope frequency $f_{o\!{f\!\!{f\!{set}}}}$ are 250 MHz and 35 MHz respectively, which are both locked to a rubidium frequency standard, and hence the frequency of its $N$th comb mode is $f_N=f_{o\!{f\!\!{f\!{set}}}}+Nf_{rep}=35$ MHz + $N\times$ 250 MHz. The double-pass configuration is applied to eliminate the alignment problem caused by AOM\cite{R23}.
 
 In principle, we use the lowest beat signal ($0\leq f_{beat} \leq f_{rep}/2=125$ MHz) for stabilizing and tuning the laser frequency. Here, we use some optical and electronic setup to eliminate other high order beat signals, as well as improving the SNR (signal to noise ratio). The combination of optical grating and the single mode fiber coupling can spatially filter the OFC whose spectrum is near the diode laser frequency. In the light path of interference before the optical grating, the power of diode laser is about 110 $\mu$W and the power of the OFC is about 7 mW. After the mode coupling of single mode fiber, the total power detected by the photodetector (FPD 310, Menlo System) is 105 $\mu$W, where 56 $\mu$W is the power of diode laser and 49 $\mu$W is the power of the OFC. With some calculation and estimation, there are about 2284 modes lines exists and the power of a single mode is about 21.4 nW.
 
\begin{figure}[h]
   \centering\includegraphics[width=12cm]{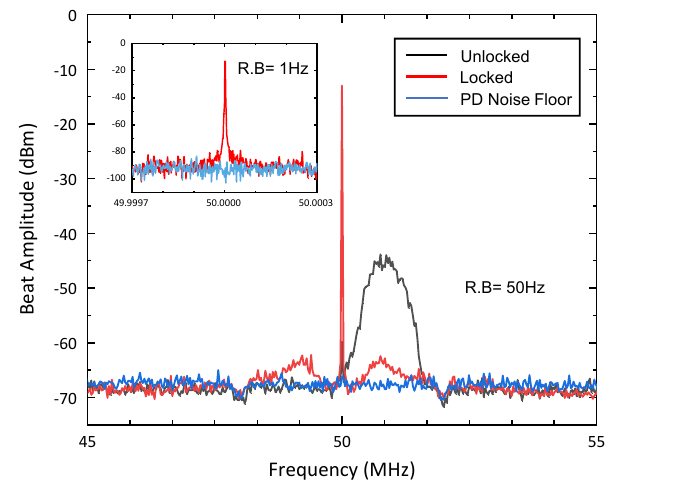}
   \caption{ 
   \label{lock}  
   The beat signal between the tunable laser and one comb mode at a resolution bandwidth of 50 Hz. The black curve is the free-running beat signal and the red curve is the beat signal as the laser is locked to the OFC. The PD noise floor is depicted as blue curve. The inset is at a resolution bandwidth of 1 Hz. }
 \end{figure}

 Then the beat signal is transferred to the electronic setup, in which it passes through a low pass filter (SLP-200+, Mini Circuits) with a cutoff at 210 MHz and is amplified by a RF amplifier. The amplified beat signal is divided into two parts. One part is used for observation with a spectrum analyzer, and the other part is transferred to the mixer, mixed with the local oscillator (LO) signal. The LO (Rigol, DG4162) is an accessible signal generator which can be triggered to sweep the frequency of the output sinusoidal signal. The output of the mixer is transferred to the fast laser locking module (Falc 110, Toptica) as the error signal. Falc 110 is a module for laser frequency stabilization tasks, such as high-bandwidth locking and linewidth reduction. This locking module provides feedback to the the laser diode current and the offset voltage of the piezo-electric transducer. No matter the feasible set point is 0 V or other fixed values, the laser frequency is adjusted by the locking module, suppressing the error signal to be an DC signal. In this way, the frequency of the beat signal is locked to an RF source. Since the laser frequency can be expressed as $f_{laser}=f_N+f_{beat}=f_{o\!{f\!\!{f\!{set}}}}+Nf_{rep}+f_{beat}$, where $f_N$ and $f_{beat}$ have been locked to the transition of Rubidium and an RF source respectively, the laser frequency is successfully stabilized. The performance of the above frequency locking is depicted in FIG.~\ref{lock}. The linewidth of the beat signal between the ECDL and one comb mode is reduced effectively by the feedback. After such locking, we can infer that the characteristics of linewidth and stability is mainly transferred from the OFC to the ECDL. 
 
 \section{Laser tuning and frequency jump}
 
 \begin{figure*}[t]
   \centering
   \includegraphics[width=12cm]{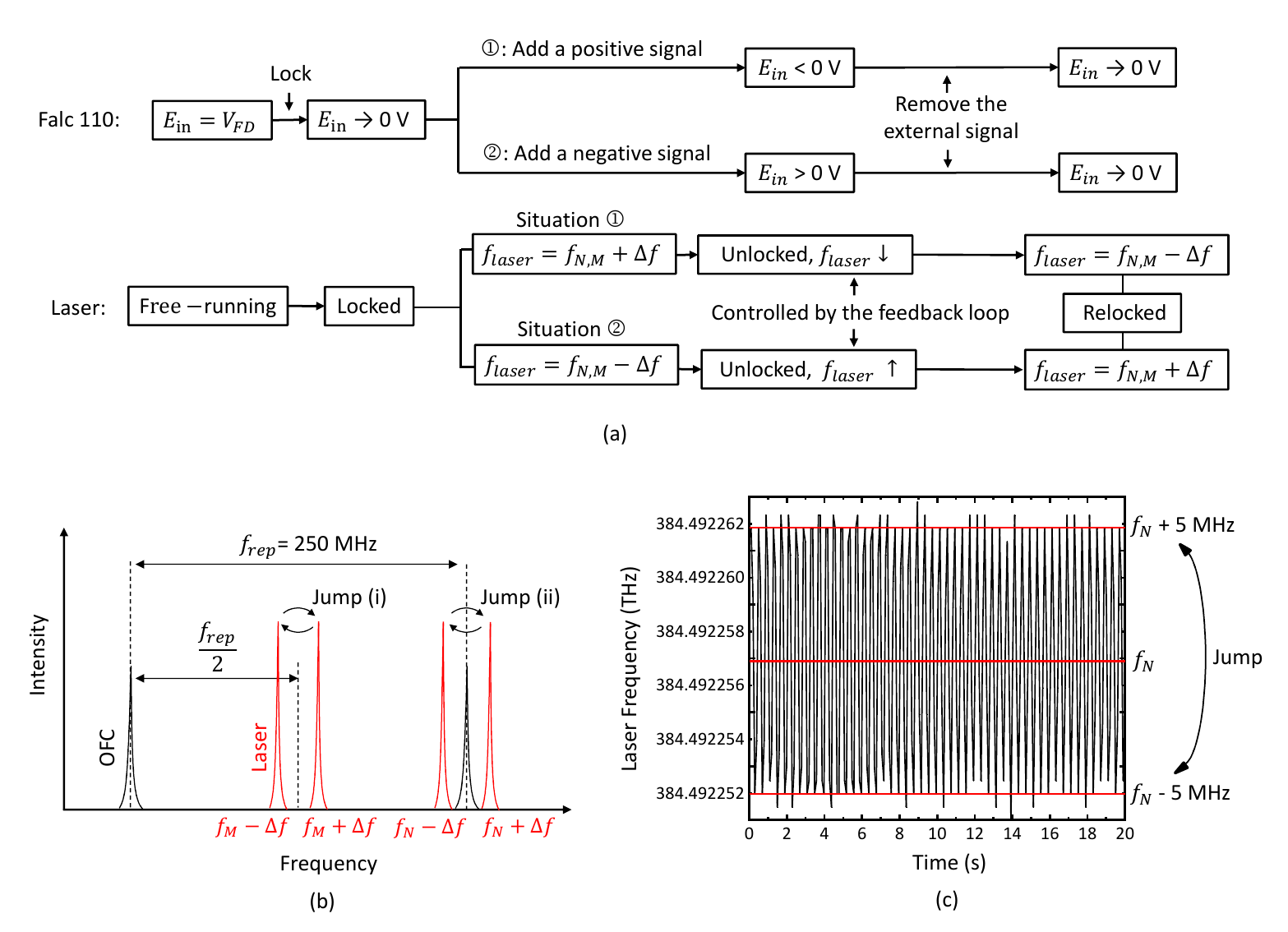}
   \caption{
   \label{jump}  
   (a) The principle of the frequency jump. $E_{in}$: the total input of the locking module; $V_{FD}$: voltage of the frequency difference between beat signal and LO signal; $f_{N,M}$: $f_{N}$ or $f_{M}$; \ding{172}: for the situation that the locked laser frequency is equal to $f_{N,M}+\varDelta f$; \ding{173}: for the situation that the locked laser frequency is equal to $f_{N,M}-\varDelta f$. (b) The schematic of the laser frequency jump in the frequency domain. Jump (i): jump over the comb mode. Jump (ii): jump over the middle position of two adjacent come modes. (c) The frequency jumps measured by the wavemeter. We add the positive pulse and the negative pulse to the error signal in cycles, driving the laser frequency to jump back and forth around its nearest comb mode.}
 \end{figure*}
 
 As the beat signal is locked to the LO, its frequency as well as the laser frequency varies with the changing of LO frequency. However, when the laser frequency is reaching the middle position of two adjacent comb modes, which means the beat signal is reaching 125 MHz in our setup, the other beat signal between the laser frequency and its next nearest neighbor comb mode is too close to the prior one, causing confusion for the locking module and then the locking module is invalid. Therefore, in our experiment, the LO signal can vary from 5 MHz, which is limited by the bandwidth of FPD, to 124.5 MHz, which is close to the ambiguity zone in the middle of two adjacent comb modes. In addition, the beat signal can't move through 0 MHz neither, since the DC error signal provides invalid information for feedback. The existence of these two ambiguity zones, 0 MHz and $f_{rep}/2$, blocks the wide sweep of the beat signal and thus hinders the laser to be tuned continuously and widely. 
 
 In order to settle the above matter, we add an external pulse signal to the error signal, driving the beat signal to jump over the ambiguity zones without suspending the feedback.  The error signal and external pulse signal are sent to the inverting input and the non-inverting input of Falc 110 respectively. Therefore, the total input of Falc 110 is the minus error signal plus the pulse signal and expressed as $E_{in}$. When $E_{in}$ is beyond or below the lock point, such as 0 V, the locking module will adjust the laser frequency, attempting to pull the input voltage to zero. The basic principle of the frequency jump is shown in FIG.~\ref{jump} (a). As the voltage of the external signal is at 0 V, $E_{in}$ is equal to the voltage of the frequency difference between beat signal and LO signal. Then the locking module works to suppress it to 0 V and the $f_{laser}$ is locked to $f_{N}\pm\varDelta f$ or  $f_{M}\pm\varDelta f$ shown in ~\ref{jump} (b), where $f_{M}$ is the middle frequency of two adjacent comb modes and $\varDelta f$ is equal to the frequency of LO signal. Take the situation \ding{172} that the laser frequency is locked to $f_{N} + \varDelta f$ or $f_{M} + \varDelta f$ for instance. As the external positive pulse comes, $E_{in}$ will be beyond the zero level completely if the voltage of the added signal is higher than the peak voltage of the original error signal. Then the module attempts to pull $E_{in}$ back to zero, driving the laser frequency to tune in an specific direction, such as being lower in our setup. Only as the external signal is removed, the locking module is likely to relock the laser frequency. If the laser frequency is pulled from $f_{N} + \varDelta f$ to the value around $f_{N} - \varDelta f$ for example, which means the beat signal is set close to its initial value, the laser frequency will be relocked successfully. 
 
 Therefore, we can drive the desired frequency jump, such as Jump (i) and Jump (ii) shown in ~\ref{jump} (b), by adding an appropriate pulse signal to the error signal with the always-activated feedback loop. Jump (i) means the laser frequency jumps over the middle position of two adjacent comb modes and Jump (ii) means the laser frequency jumps over its nearest comb mode. The polarity of the pulse signal determines the direction of the frequency jump. A positive pulse, for example, drives the laser frequency to jump to to the symmetrical position of a lower frequency in our setup. Here, we add the positive pulse and the negative pulse in cycles to drive the laser frequency to jump back and forth around its nearest comb mode, which corresponds to the Jump (ii). The width of the added pulse is 50 us and their amplitude is 600 mV and -600 mV respectively before passing an 20 dB attenuator, which is used to reduce the intrinsic electronic noise. About a hundred frequency jumps measured by the wavemeter are shown in FIG.~\ref{jump} (c). The exposure time of the wavemeter is 100 ms and thus the interval between two frequency jumps is set as 200ms to distinguish each frequency jump. Due to the measurement resolution of the wavemeter, some data is not critically at the red line shown in FIG.~\ref{jump} (c). However, we can verify that the laser frequency is successfully relocked after each frequency jump by the observation with the spectrum analyzer in FIG.1. We test more than one thousand times uninterruptedly to prove that it is robust and no error occurs. The rise time of the frequency jump is on the order of ten microseconds, which can be inferred from the width of the added pulse. It is much faster than 2.5 ms shown in Ref.~\onlinecite{R18}, since it is mainly controlled by the diode current instead of the PZT.
 
 \begin{figure}[h]
   \centering
   \includegraphics[width=12cm]{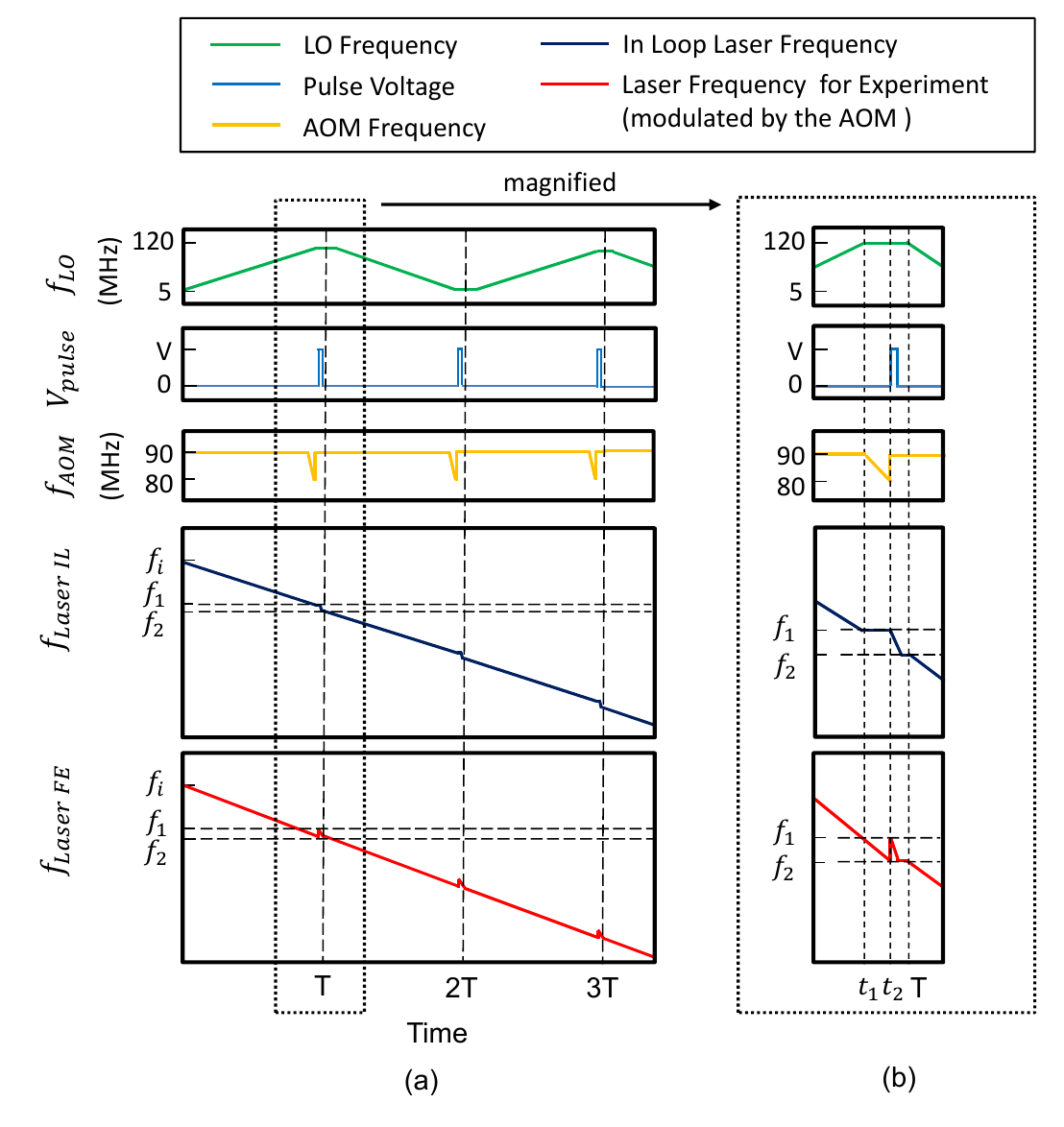}
   \caption{ 
     \label{time}
     (a) The time sequence to tune the laser continuously until the mode hopping happens. The laser frequency is tuned $f_{rep}/2$ = 125 MHz in each cycle. (b) Enlarged drawing of the dotted box in figure (a). The laser frequency tuning is precisely linear shown in the period from $0$ to $t_2$ and there is none of missing point by the compensation with the AOM. $f_{LO}$: LO frequency; $V_{pulse}$: pulse voltage; $f_{AOM}$: AOM frequency; $f_{Laser IL}:$ frequency of the laser in the locking loop, $f_{Laser FE}$: frequency of the laser for experiment, which is modulated by an AOM. }
 \end{figure}
 
 The added pulse signal is provided by the analog card, which is also used to synchronize the time sequence. In our setup, it triggers the sweep of LO after each jump and controls the AOM to compensate the gap of each frequency jump via a VCO. The time sequence of the laser tuning is depicted in FIG.~\ref{time}. In the beginning, the laser frequency is locked at the frequency which is 5 MHz less than the frequency of its nearest comb mode which means the beat signal locked on the LO signal is 5 MHz as well. As the LO frequency sweeps from 5 MHz to 120 MHz linearly, the laser frequency is tuned accordingly from $f_i$ (initial frequency) to $f_1=f_i-115$ MHz from 0 to $t_1$ shown in FIG. 4. When the sweep of LO frequency stops at time $t_1$, the AOM modulates the laser frequency from $f_1=f_i-115$ MHz to $f_2=f_i-125$ MHz in the period from $t_1$ to $t_2$ at the same rate as the previous sweep rate of the LO frequency. Then we reset the AOM frequency to the its initial value and add the pulse signal to drive the beat signal to jump at $t_2$. After resetting the AOM and driving the frequency jump mentioned above, the laser frequency is still $f_2=f_i-125$ MHz which ensures that there is none of missing points in one cycle and the laser tuning from $f_i$ to $f_i$ -125 MHz is precisely linear in the period from 0 to $t_2$. Then we can repeat above steps to tune the laser frequency until the mode hopping happens.
 The tuning range of the laser frequency from $t_2$ to T is the same as the rang from $t_1$ to $t_2$, but less precise for the experiment, which means it is useless. Thus the time from $t_2$ to T to reset the AOM and drive the frequency jump should be condensed to improve the average tuning rate in the practical experiment. In addition, the time nodes, $t_2$ and T, are known exactly. Therefore, we can dispose the data from $t_2$ to T accurately and just leave the precisely linear laser tuning behind. With the above data processing, the continuous, wide and precisely linear laser tuning can be realized.
 
 \section{Performance}
 
\begin{figure}[h!]
   \centering
   \includegraphics[width=10cm]{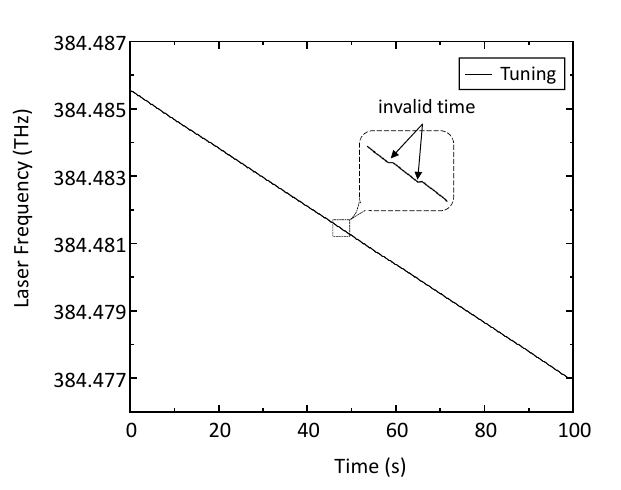}
   \caption{
   \label{result_1}  
   Laser tuning for examination. Between the short broken lines which correspond to the invalid time, the laser is tuned linearly.}
 \end{figure}

 With the above time sequence, we set the sweep speed of the LO frequency and the AOM frequency to 100 MHz/s to test the laser tuning and the result is demonstrated in FIG.~\ref{result_1}. The tuning range is nearly 10 GHz, limited by the mode hopping of the ECDL. The time during which we reset the AOM and drive the beat signal to jump over the ambiguity zones is depicted as short broken lines, due to the sampling rate of the wavemeter. It is demonstrated in the previous section that these short broken lines corresponds to the invalid time and they can be disposed accurately. The laser is tuned linearly and smoothly as expected between these short broken lines. Thus the feasibility of this scheme is proved.
 
\begin{figure}[h!]
   \centering
   \includegraphics[width=10cm]{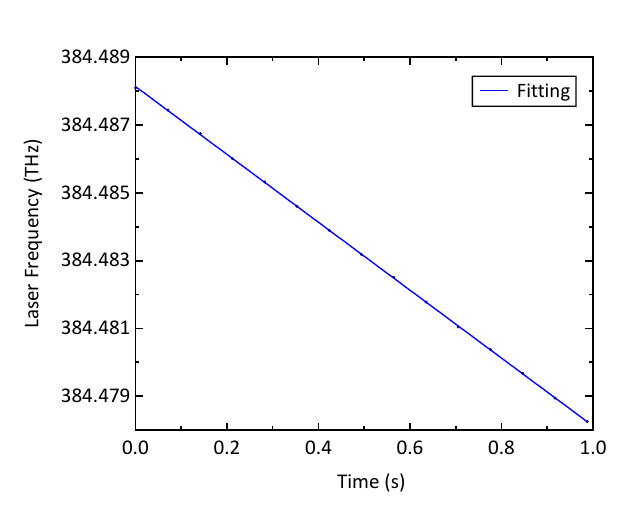}
   \caption{
   \label{result_2}  
   Laser tuning in experiments after disposing the invalid data. The data is analyzed by linear fitting and the fitting speed is equal to the sweep rate, showing that no error of frequency jump occurs in the whole tuning process.}
 \end{figure}

 In the practical experiment, we set the sweep speed of the LO frequency and the AOM frequency to 10 GHz/s and reserve 5 milliseconds to drive the beat signal and reset the AOM. The period of laser tuning T is 125 MHz / (10 GHz/s) + 5 ms = 17.5 ms which means there is invalid time of 5 ms in each 17.5 ms for each cycle and the average tuning rate is 125 MHz/ 17.5 ms = 7.14 GHz/s. With the above calculation, the measured data of laser tuning after accurately disposing the invalid time is shown in FIG.~\ref{result_2}. The sample rate of the wavemeter limited by its exposure time makes the amount of the data not enough to show details of the tuning process. Therefore, we apply linear fitting to analyze the data. The fitting speed of laser tuning is 10 ($\pm$ 0.016) GHz/s which corresponds to the sweep speed and the fitting result shows a good linearity, proving that the laser tunes as expected and there is none of failure of any frequency jump during the tuning process. It is worth mentioning that the tuning rate in our set up is limited mechanically by the response speed of the PZT.
 
\begin{figure}[h!]
   \centering
   \includegraphics[width=10cm]{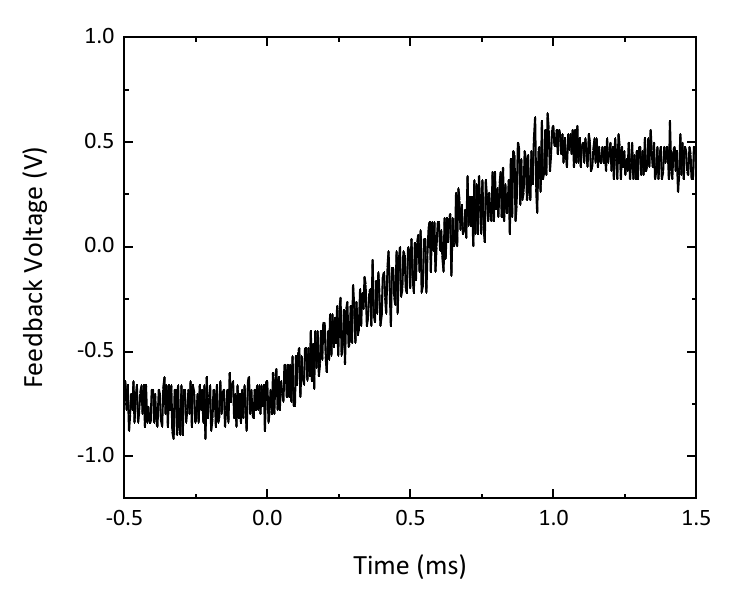}
   \caption{
   \label{sweep}  
   The detection of the frequency sweep. The rise time of the feedback signal corresponds to the setting sweep time, 1 ms.}
 \end{figure}

 To improve the tuning rate and prove the universality of this method, we turn off the feedback of PZT and tune the frequency-stabilized laser only with the feedback of the diode current. In this situation, the tuning between two dead zones can be completed within 1 ms, which is limited by the radio-frequency source, not the method or laser itself. The voltage of the feedback shown in Fig. ~\ref{sweep} is observed with an oscilloscope to prove the frequency is tuned exactly within 1 ms. 
 
\begin{figure}[h!]
   \centering
   \includegraphics[width=10cm]{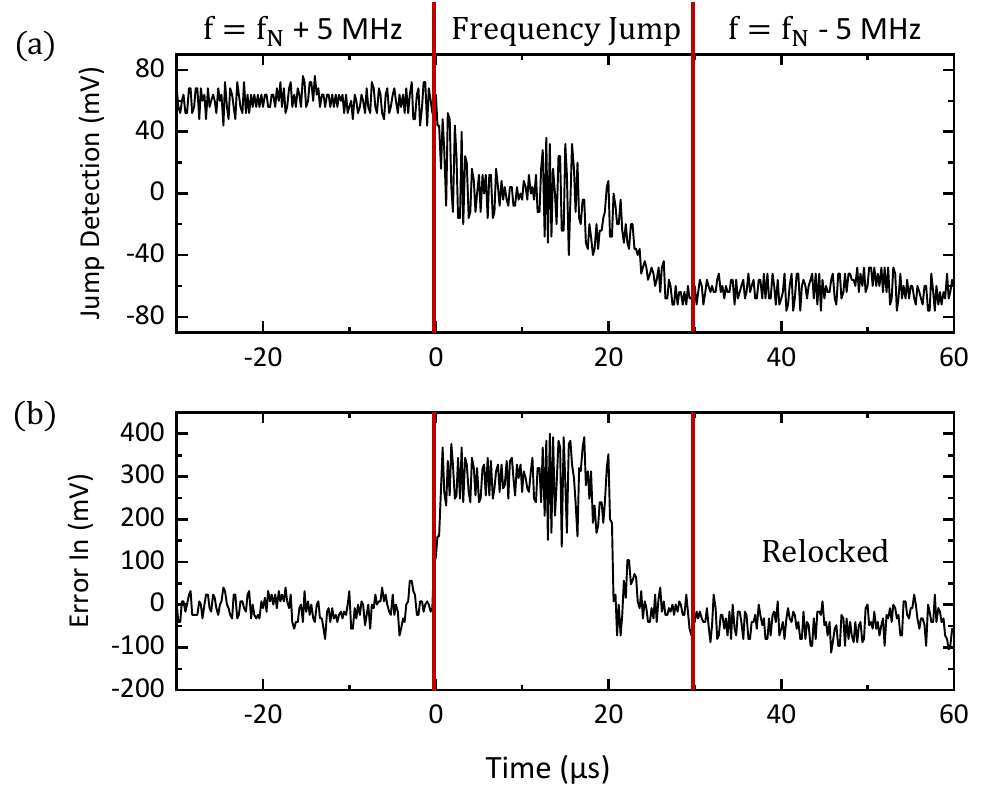}
   \caption{
   \label{jump time}  
    The detection of the frequency jump. (a) The output signal of the mixer which is fed with the beat signal and a 5 MHz RF signal. (b) The total input of FALC 110, which equals minus the error signal plus the pulse signal.}
 \end{figure}

 The setting width of the pulse signal to drive the frequency jump is 20 $\mu$s in this situation and the time for the frequency jump is measured here. The beat signal and another 5 MHz RF signal are put into a mixer and its output is detected with an oscillator. Since the beat signal jump from +5 MHz to -5 MHz for example, the output of the mixer should vary from cos(5 MHz) cos(5 MHz) = 1/2 + cos(10 MHz) to cos (5 MHz + $\pi $) cos(5 MHz)= -1/2 - cos(10 MHz). Connecting the mixer output with a 1.9 MHz low pass filter to remove the 10 MHz signal, the output indeed jumps from a positive voltage to a negative voltage shown in the top of Fig. ~\ref{jump time}, from which we can see that the time for frequency jump is about 30 $\mu$s. The bottom of Fig. ~\ref{jump time}  shows the total input of FALC 110, which equals the minus error signal plus the pulse signal. The width of the pulse signal added to the error signal is 20 $\mu$s corresponding to the setting.
 
\begin{figure}[h!]
  \centering
  \includegraphics[width=10cm]{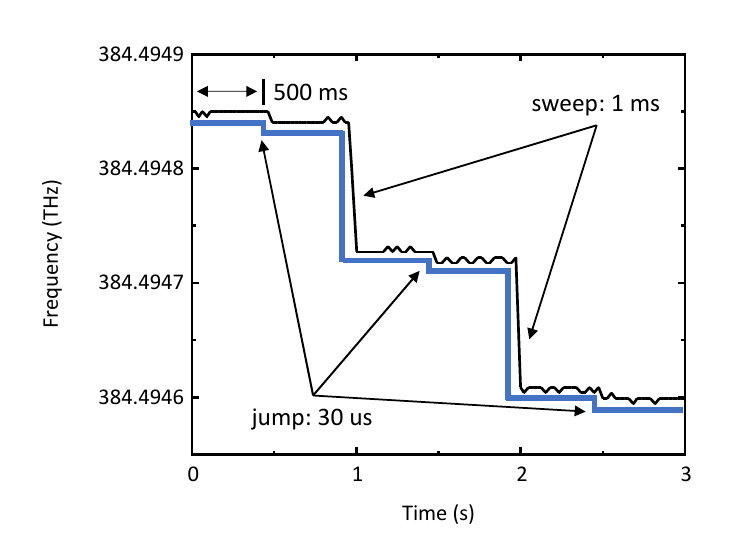}
  \caption{
  \label{tuning with current}  
  Tuning result of combining the frequency sweep and jump as the ECDL is only controlled with the feedback of diode current. A waiting time of 500 ms is added to distinguish each step.}
\end{figure}

 If the frequency jump and sweep are combined in this situation, the average tuning rate is (115 MHz + 10 MHz)/(1 ms + 30 $\mu$s) = 121.36 GHz/s. Unfortunately, the tuning range without the mode hoping is rather small in this situation, since the PZT inside ECDL does not work. Meanwhile, the ideal fastest measurement speed of the wavemeter is 500 Hz, corresponding to a resolution of 2 ms. If we tune the laser over 300 MHz at a rate of 100 GHz/s for example, it takes only 3 ms which cannot be observed clearly by the wavemeter. Here, we add the waiting time of 500 ms between the frequency jump and frequency sweep to distinguish each step. The tuning result is shown in Fig. ~\ref{tuning with current} and a blue line is used to show each process more clearly.

 \section{Conclusion and outlook}
 
 In conclusion, we demonstrate a wide, continuous and precise laser tuning method applied to an ECDL with some readily available electronics. In our scheme, the laser is stabilized on the OFC and its frequency is tuned at nearly 7 GHz/s without suspending the feedback loop. The always-activated feedback loop reduces the complexity of reengaging the servo loop compared with the previous reports\cite{R18,R19} and its spirit can be referenced in other applications in which the locking status faces challenges around some specific areas during the scanning. Since an AOM is used to compensate the frequency gap and the time nodes of the frequency jump are determined exactly, we can get the continuous and precisely linear tuning data after the processing mentioned above. 
 
 It is worth noting that the above tuning rate is mechanically limited by the PZT inside the ECDL and thus we turn off the feedback of PZT and control the ECDL only with the feedback of diode current. In this situation, the average tuning speed can exceed 100 GHz/s though the tuning range without mode hopping is rather small. We believe this scheme can be applied to other types of lasers, such as the distributed feedback laser whose frequency can be only controlled by the current. Meanwhile, since there is none of adjustment to the parameters of the OFC during the whole process, we can lock and tune multiple lasers independently on one OFC. Thanks to the resource-efficient configuration and its simplicity of completion, this scheme can be referenced and find wide applications in AMO experiments.
 
 \section*{acknowledgments}
 \vspace{-1em} 
   This work was supported by Innovation Program for Quantum Science and Technology (No. 2021ZD0301200), National Natural Science Foundation of China (11804330,11821404) and the Fundamental Research Funds for the Central Universities.
 \\
 
 \section*{AUTHOR DECLARATIONS}
 \vspace{-1em} 
  \subsection*{Conflict of Interest} 
  \vspace{-1em} 
    The authors have no conflicts to disclose.
 
 \section*{DATA AVAILABILITY} 
 \vspace{-1em} 
 The data that support the findings of this study are available from the corresponding author upon reasonable request.
  
 \vspace{-1em} 
 \bibliography{Manuscript.bib}

\begin{thebibliography}{32}%
\makeatletter
\providecommand \@ifxundefined [1]{%
 \@ifx{#1\undefined}
}%
\providecommand \@ifnum [1]{%
 \ifnum #1\expandafter \@firstoftwo
 \else \expandafter \@secondoftwo
 \fi
}%
\providecommand \@ifx [1]{%
 \ifx #1\expandafter \@firstoftwo
 \else \expandafter \@secondoftwo
 \fi
}%
\providecommand \natexlab [1]{#1}%
\providecommand \enquote  [1]{``#1''}%
\providecommand \bibnamefont  [1]{#1}%
\providecommand \bibfnamefont [1]{#1}%
\providecommand \citenamefont [1]{#1}%
\providecommand \href@noop [0]{\@secondoftwo}%
\providecommand \href [0]{\begingroup \@sanitize@url \@href}%
\providecommand \@href[1]{\@@startlink{#1}\@@href}%
\providecommand \@@href[1]{\endgroup#1\@@endlink}%
\providecommand \@sanitize@url [0]{\catcode `\\12\catcode `\$12\catcode
  `\&12\catcode `\#12\catcode `\^12\catcode `\_12\catcode `\%12\relax}%
\providecommand \@@startlink[1]{}%
\providecommand \@@endlink[0]{}%
\providecommand \url  [0]{\begingroup\@sanitize@url \@url }%
\providecommand \@url [1]{\endgroup\@href {#1}{\urlprefix }}%
\providecommand \urlprefix  [0]{URL }%
\providecommand \Eprint [0]{\href }%
\providecommand \doibase [0]{http://dx.doi.org/}%
\providecommand \selectlanguage [0]{\@gobble}%
\providecommand \bibinfo  [0]{\@secondoftwo}%
\providecommand \bibfield  [0]{\@secondoftwo}%
\providecommand \translation [1]{[#1]}%
\providecommand \BibitemOpen [0]{}%
\providecommand \bibitemStop [0]{}%
\providecommand \bibitemNoStop [0]{.\EOS\space}%
\providecommand \EOS [0]{\spacefactor3000\relax}%
\providecommand \BibitemShut  [1]{\csname bibitem#1\endcsname}%
\let\auto@bib@innerbib\@empty
%</preamble>
\bibitem [{\citenamefont {Juncar}\ \emph {et~al.}(1981)\citenamefont {Juncar},
  \citenamefont {Pinard}, \citenamefont {Hamon},\ and\ \citenamefont
  {Chartier}}]{X1}%
  \BibitemOpen
  \bibfield  {author} {\bibinfo {author} {\bibfnamefont {P.}~\bibnamefont
  {Juncar}}, \bibinfo {author} {\bibfnamefont {J.}~\bibnamefont {Pinard}},
  \bibinfo {author} {\bibfnamefont {J.}~\bibnamefont {Hamon}}, \ and\ \bibinfo
  {author} {\bibfnamefont {A.}~\bibnamefont {Chartier}},\ }\href {\doibase
  10.1088/0026-1394/17/3/001} {\bibfield  {journal} {\bibinfo  {journal}
  {Metrologia}\ }\textbf {\bibinfo {volume} {17}},\ \bibinfo {pages} {77}
  (\bibinfo {year} {1981})}\BibitemShut {NoStop}%
\bibitem [{\citenamefont {Zheng}\ \emph {et~al.}(2021)\citenamefont {Zheng},
  \citenamefont {Cui}, \citenamefont {Ai}, \citenamefont {Qian}, \citenamefont
  {Cao}, \citenamefont {Huang}, \citenamefont {Jia}, \citenamefont {Li},\ and\
  \citenamefont {Guo}}]{R2}%
  \BibitemOpen
  \bibfield  {author} {\bibinfo {author} {\bibfnamefont {Y.-X.}\ \bibnamefont
  {Zheng}}, \bibinfo {author} {\bibfnamefont {J.-M.}\ \bibnamefont {Cui}},
  \bibinfo {author} {\bibfnamefont {M.-Z.}\ \bibnamefont {Ai}}, \bibinfo
  {author} {\bibfnamefont {Z.-h.}\ \bibnamefont {Qian}}, \bibinfo {author}
  {\bibfnamefont {H.}~\bibnamefont {Cao}}, \bibinfo {author} {\bibfnamefont
  {Y.-F.}\ \bibnamefont {Huang}}, \bibinfo {author} {\bibfnamefont {X.-J.}\
  \bibnamefont {Jia}}, \bibinfo {author} {\bibfnamefont {C.-F.}\ \bibnamefont
  {Li}}, \ and\ \bibinfo {author} {\bibfnamefont {G.-C.}\ \bibnamefont {Guo}},\
  }\href {\doibase 10.1364/OE.433207} {\bibfield  {journal} {\bibinfo
  {journal} {Opt. Express}\ }\textbf {\bibinfo {volume} {29}},\ \bibinfo
  {pages} {24674} (\bibinfo {year} {2021})}\BibitemShut {NoStop}%
\bibitem [{\citenamefont {Englert}\ \emph {et~al.}(2011)\citenamefont
  {Englert}, \citenamefont {Mielenz}, \citenamefont {Sommer}, \citenamefont
  {Bayerl}, \citenamefont {Motsch}, \citenamefont {Pinkse}, \citenamefont
  {Rempe},\ and\ \citenamefont {Zeppenfeld}}]{Z2}%
  \BibitemOpen
  \bibfield  {author} {\bibinfo {author} {\bibfnamefont {B.~G.~U.}\
  \bibnamefont {Englert}}, \bibinfo {author} {\bibfnamefont {M.}~\bibnamefont
  {Mielenz}}, \bibinfo {author} {\bibfnamefont {C.}~\bibnamefont {Sommer}},
  \bibinfo {author} {\bibfnamefont {J.}~\bibnamefont {Bayerl}}, \bibinfo
  {author} {\bibfnamefont {M.}~\bibnamefont {Motsch}}, \bibinfo {author}
  {\bibfnamefont {P.~W.~H.}\ \bibnamefont {Pinkse}}, \bibinfo {author}
  {\bibfnamefont {G.}~\bibnamefont {Rempe}}, \ and\ \bibinfo {author}
  {\bibfnamefont {M.}~\bibnamefont {Zeppenfeld}},\ }\href {\doibase
  10.1103/PhysRevLett.107.263003} {\bibfield  {journal} {\bibinfo  {journal}
  {Phys. Rev. Lett.}\ }\textbf {\bibinfo {volume} {107}},\ \bibinfo {pages}
  {263003} (\bibinfo {year} {2011})}\BibitemShut {NoStop}%
\bibitem [{\citenamefont {Glöckner}\ \emph {et~al.}(2015)\citenamefont
  {Glöckner}, \citenamefont {Prehn}, \citenamefont {Englert}, \citenamefont
  {Rempe},\ and\ \citenamefont {Zeppenfeld}}]{Z3}%
  \BibitemOpen
  \bibfield  {author} {\bibinfo {author} {\bibfnamefont {R.}~\bibnamefont
  {Glöckner}}, \bibinfo {author} {\bibfnamefont {A.}~\bibnamefont {Prehn}},
  \bibinfo {author} {\bibfnamefont {B.~G.~U.}\ \bibnamefont {Englert}},
  \bibinfo {author} {\bibfnamefont {G.}~\bibnamefont {Rempe}}, \ and\ \bibinfo
  {author} {\bibfnamefont {M.}~\bibnamefont {Zeppenfeld}},\ }\href {\doibase
  10.1103/PhysRevLett.115.233001} {\bibfield  {journal} {\bibinfo  {journal}
  {Phys. Rev. Lett.}\ }\textbf {\bibinfo {volume} {115}},\ \bibinfo {pages}
  {233001} (\bibinfo {year} {2015})}\BibitemShut {NoStop}%
\bibitem [{\citenamefont {Uruñuela}\ \emph {et~al.}(2020)\citenamefont
  {Uruñuela}, \citenamefont {Alt}, \citenamefont {Keiler}, \citenamefont
  {Meschede}, \citenamefont {Pandey}, \citenamefont {Pfeifer},\ and\
  \citenamefont {Macha}}]{Z4}%
  \BibitemOpen
  \bibfield  {author} {\bibinfo {author} {\bibfnamefont {E.}~\bibnamefont
  {Uruñuela}}, \bibinfo {author} {\bibfnamefont {W.}~\bibnamefont {Alt}},
  \bibinfo {author} {\bibfnamefont {E.}~\bibnamefont {Keiler}}, \bibinfo
  {author} {\bibfnamefont {D.}~\bibnamefont {Meschede}}, \bibinfo {author}
  {\bibfnamefont {D.}~\bibnamefont {Pandey}}, \bibinfo {author} {\bibfnamefont
  {H.}~\bibnamefont {Pfeifer}}, \ and\ \bibinfo {author} {\bibfnamefont
  {T.}~\bibnamefont {Macha}},\ }\href {\doibase 10.1103/PhysRevA.101.023415}
  {\bibfield  {journal} {\bibinfo  {journal} {Phys. Rev. A}\ }\textbf {\bibinfo
  {volume} {101}},\ \bibinfo {pages} {023415} (\bibinfo {year}
  {2020})}\BibitemShut {NoStop}%
\bibitem [{\citenamefont {Holzwarth}\ \emph {et~al.}(2000)\citenamefont
  {Holzwarth}, \citenamefont {Udem}, \citenamefont {H\"ansch}, \citenamefont
  {Knight}, \citenamefont {Wadsworth},\ and\ \citenamefont {Russell}}]{R5}%
  \BibitemOpen
  \bibfield  {author} {\bibinfo {author} {\bibfnamefont {R.}~\bibnamefont
  {Holzwarth}}, \bibinfo {author} {\bibfnamefont {T.}~\bibnamefont {Udem}},
  \bibinfo {author} {\bibfnamefont {T.~W.}\ \bibnamefont {H\"ansch}}, \bibinfo
  {author} {\bibfnamefont {J.~C.}\ \bibnamefont {Knight}}, \bibinfo {author}
  {\bibfnamefont {W.~J.}\ \bibnamefont {Wadsworth}}, \ and\ \bibinfo {author}
  {\bibfnamefont {P.~S.~J.}\ \bibnamefont {Russell}},\ }\href {\doibase
  10.1103/PhysRevLett.85.2264} {\bibfield  {journal} {\bibinfo  {journal}
  {Phys. Rev. Lett.}\ }\textbf {\bibinfo {volume} {85}},\ \bibinfo {pages}
  {2264} (\bibinfo {year} {2000})}\BibitemShut {NoStop}%
\bibitem [{\citenamefont {Ludlow}\ \emph {et~al.}(2015)\citenamefont {Ludlow},
  \citenamefont {Boyd}, \citenamefont {Ye}, \citenamefont {Peik},\ and\
  \citenamefont {Schmidt}}]{R6}%
  \BibitemOpen
  \bibfield  {author} {\bibinfo {author} {\bibfnamefont {A.~D.}\ \bibnamefont
  {Ludlow}}, \bibinfo {author} {\bibfnamefont {M.~M.}\ \bibnamefont {Boyd}},
  \bibinfo {author} {\bibfnamefont {J.}~\bibnamefont {Ye}}, \bibinfo {author}
  {\bibfnamefont {E.}~\bibnamefont {Peik}}, \ and\ \bibinfo {author}
  {\bibfnamefont {P.~O.}\ \bibnamefont {Schmidt}},\ }\href {\doibase
  10.1103/RevModPhys.87.637} {\bibfield  {journal} {\bibinfo  {journal} {Rev.
  Mod. Phys.}\ }\textbf {\bibinfo {volume} {87}},\ \bibinfo {pages} {637}
  (\bibinfo {year} {2015})}\BibitemShut {NoStop}%
\bibitem [{\citenamefont {Hisai}\ \emph {et~al.}(2019)\citenamefont {Hisai},
  \citenamefont {Akamatsu}, \citenamefont {Kobayashi}, \citenamefont {Okubo},
  \citenamefont {Inaba}, \citenamefont {Hosaka}, \citenamefont {Yasuda},\ and\
  \citenamefont {Hong}}]{R7}%
  \BibitemOpen
  \bibfield  {author} {\bibinfo {author} {\bibfnamefont {Y.}~\bibnamefont
  {Hisai}}, \bibinfo {author} {\bibfnamefont {D.}~\bibnamefont {Akamatsu}},
  \bibinfo {author} {\bibfnamefont {T.}~\bibnamefont {Kobayashi}}, \bibinfo
  {author} {\bibfnamefont {S.}~\bibnamefont {Okubo}}, \bibinfo {author}
  {\bibfnamefont {H.}~\bibnamefont {Inaba}}, \bibinfo {author} {\bibfnamefont
  {K.}~\bibnamefont {Hosaka}}, \bibinfo {author} {\bibfnamefont
  {M.}~\bibnamefont {Yasuda}}, \ and\ \bibinfo {author} {\bibfnamefont {F.-L.}\
  \bibnamefont {Hong}},\ }\href {\doibase 10.1364/OE.27.006404} {\bibfield
  {journal} {\bibinfo  {journal} {Opt. Express}\ }\textbf {\bibinfo {volume}
  {27}},\ \bibinfo {pages} {6404} (\bibinfo {year} {2019})}\BibitemShut
  {NoStop}%
\bibitem [{\citenamefont {Kwee}\ \emph {et~al.}(2012)\citenamefont {Kwee},
  \citenamefont {Bogan}, \citenamefont {Danzmann}, \citenamefont {Frede},
  \citenamefont {Kim}, \citenamefont {King}, \citenamefont {P\"{o}ld},
  \citenamefont {Puncken}, \citenamefont {Savage}, \citenamefont {Seifert},
  \citenamefont {Wessels}, \citenamefont {Winkelmann},\ and\ \citenamefont
  {Willke}}]{R8}%
  \BibitemOpen
  \bibfield  {author} {\bibinfo {author} {\bibfnamefont {P.}~\bibnamefont
  {Kwee}}, \bibinfo {author} {\bibfnamefont {C.}~\bibnamefont {Bogan}},
  \bibinfo {author} {\bibfnamefont {K.}~\bibnamefont {Danzmann}}, \bibinfo
  {author} {\bibfnamefont {M.}~\bibnamefont {Frede}}, \bibinfo {author}
  {\bibfnamefont {H.}~\bibnamefont {Kim}}, \bibinfo {author} {\bibfnamefont
  {P.}~\bibnamefont {King}}, \bibinfo {author} {\bibfnamefont {J.}~\bibnamefont
  {P\"{o}ld}}, \bibinfo {author} {\bibfnamefont {O.}~\bibnamefont {Puncken}},
  \bibinfo {author} {\bibfnamefont {R.~L.}\ \bibnamefont {Savage}}, \bibinfo
  {author} {\bibfnamefont {F.}~\bibnamefont {Seifert}}, \bibinfo {author}
  {\bibfnamefont {P.}~\bibnamefont {Wessels}}, \bibinfo {author} {\bibfnamefont
  {L.}~\bibnamefont {Winkelmann}}, \ and\ \bibinfo {author} {\bibfnamefont
  {B.}~\bibnamefont {Willke}},\ }\href {\doibase 10.1364/OE.20.010617}
  {\bibfield  {journal} {\bibinfo  {journal} {Opt. Express}\ }\textbf {\bibinfo
  {volume} {20}},\ \bibinfo {pages} {10617} (\bibinfo {year}
  {2012})}\BibitemShut {NoStop}%
\bibitem [{\citenamefont {Black}(2000)}]{R9}%
  \BibitemOpen
  \bibfield  {author} {\bibinfo {author} {\bibfnamefont {E.~D.}\ \bibnamefont
  {Black}},\ }\href {\doibase 10.1119/1.1286663} {\bibfield  {journal}
  {\bibinfo  {journal} {Am. J. Phys.}\ }\textbf {\bibinfo {volume} {69}},\
  \bibinfo {pages} {79} (\bibinfo {year} {2000})}\BibitemShut {NoStop}%
\bibitem [{\citenamefont {Gallego}\ \emph {et~al.}(2016)\citenamefont
  {Gallego}, \citenamefont {Ghosh}, \citenamefont {Alavi}, \citenamefont {Alt},
  \citenamefont {Martinez-Dorantes}, \citenamefont {Meschede},\ and\
  \citenamefont {Ratschbacher}}]{Z5}%
  \BibitemOpen
  \bibfield  {author} {\bibinfo {author} {\bibfnamefont {J.}~\bibnamefont
  {Gallego}}, \bibinfo {author} {\bibfnamefont {S.}~\bibnamefont {Ghosh}},
  \bibinfo {author} {\bibfnamefont {S.~K.}\ \bibnamefont {Alavi}}, \bibinfo
  {author} {\bibfnamefont {W.}~\bibnamefont {Alt}}, \bibinfo {author}
  {\bibfnamefont {M.}~\bibnamefont {Martinez-Dorantes}}, \bibinfo {author}
  {\bibfnamefont {D.}~\bibnamefont {Meschede}}, \ and\ \bibinfo {author}
  {\bibfnamefont {L.}~\bibnamefont {Ratschbacher}},\ }\href {\doibase
  10.1007/s00340-015-6281-z} {\bibfield  {journal} {\bibinfo  {journal} {Appl.
  Phys. B}\ }\textbf {\bibinfo {volume} {122}},\ \bibinfo {pages} {47}
  (\bibinfo {year} {2016})}\BibitemShut {NoStop}%
\bibitem [{\citenamefont {Saavedra}\ \emph {et~al.}(2021)\citenamefont
  {Saavedra}, \citenamefont {Pandey}, \citenamefont {Alt}, \citenamefont
  {Pfeifer},\ and\ \citenamefont {Meschede}}]{Z6}%
  \BibitemOpen
  \bibfield  {author} {\bibinfo {author} {\bibfnamefont {C.}~\bibnamefont
  {Saavedra}}, \bibinfo {author} {\bibfnamefont {D.}~\bibnamefont {Pandey}},
  \bibinfo {author} {\bibfnamefont {W.}~\bibnamefont {Alt}}, \bibinfo {author}
  {\bibfnamefont {H.}~\bibnamefont {Pfeifer}}, \ and\ \bibinfo {author}
  {\bibfnamefont {D.}~\bibnamefont {Meschede}},\ }\href {\doibase
  10.1364/OE.412273} {\bibfield  {journal} {\bibinfo  {journal} {Opt. Express}\
  }\textbf {\bibinfo {volume} {29}},\ \bibinfo {pages} {974} (\bibinfo {year}
  {2021})}\BibitemShut {NoStop}%
\bibitem [{\citenamefont {Tonyushkin}\ \emph {et~al.}(2007)\citenamefont
  {Tonyushkin}, \citenamefont {Light},\ and\ \citenamefont {Di~Rosa}}]{Z7}%
  \BibitemOpen
  \bibfield  {author} {\bibinfo {author} {\bibfnamefont {A.~A.}\ \bibnamefont
  {Tonyushkin}}, \bibinfo {author} {\bibfnamefont {A.~D.}\ \bibnamefont
  {Light}}, \ and\ \bibinfo {author} {\bibfnamefont {M.~D.}\ \bibnamefont
  {Di~Rosa}},\ }\href {\doibase 10.1063/1.2818773} {\bibfield  {journal}
  {\bibinfo  {journal} {Rev. Sci. Instrum.}\ }\textbf {\bibinfo {volume}
  {78}},\ \bibinfo {pages} {123103} (\bibinfo {year} {2007})}\BibitemShut
  {NoStop}%
\bibitem [{\citenamefont {Zhen}\ \emph {et~al.}(2008)\citenamefont {Zhen},
  \citenamefont {Ye}, \citenamefont {Liu}, \citenamefont {Zhu}, \citenamefont
  {Li}, \citenamefont {Lu},\ and\ \citenamefont {Wang}}]{Z1}%
  \BibitemOpen
  \bibfield  {author} {\bibinfo {author} {\bibfnamefont {H.}~\bibnamefont
  {Zhen}}, \bibinfo {author} {\bibfnamefont {H.}~\bibnamefont {Ye}}, \bibinfo
  {author} {\bibfnamefont {X.}~\bibnamefont {Liu}}, \bibinfo {author}
  {\bibfnamefont {D.}~\bibnamefont {Zhu}}, \bibinfo {author} {\bibfnamefont
  {H.}~\bibnamefont {Li}}, \bibinfo {author} {\bibfnamefont {Y.}~\bibnamefont
  {Lu}}, \ and\ \bibinfo {author} {\bibfnamefont {Q.}~\bibnamefont {Wang}},\
  }\href {\doibase 10.1364/oe.16.009595} {\bibfield  {journal} {\bibinfo
  {journal} {Opt. Express}\ }\textbf {\bibinfo {volume} {16}},\ \bibinfo
  {pages} {9595} (\bibinfo {year} {2008})}\BibitemShut {NoStop}%
\bibitem [{\citenamefont {Thorpe}\ \emph {et~al.}(2008)\citenamefont {Thorpe},
  \citenamefont {Numata},\ and\ \citenamefont {Livas}}]{R10}%
  \BibitemOpen
  \bibfield  {author} {\bibinfo {author} {\bibfnamefont {J.~I.}\ \bibnamefont
  {Thorpe}}, \bibinfo {author} {\bibfnamefont {K.}~\bibnamefont {Numata}}, \
  and\ \bibinfo {author} {\bibfnamefont {J.}~\bibnamefont {Livas}},\ }\href
  {\doibase 10.1364/OE.16.015980} {\bibfield  {journal} {\bibinfo  {journal}
  {Opt. Express}\ }\textbf {\bibinfo {volume} {16}},\ \bibinfo {pages} {15980}
  (\bibinfo {year} {2008})}\BibitemShut {NoStop}%
\bibitem [{\citenamefont {Diddams~Scott}\ \emph {et~al.}(2020)\citenamefont
  {Diddams~Scott}, \citenamefont {Vahala},\ and\ \citenamefont {Udem}}]{R11}%
  \BibitemOpen
  \bibfield  {author} {\bibinfo {author} {\bibfnamefont {A.}~\bibnamefont
  {Diddams~Scott}}, \bibinfo {author} {\bibfnamefont {K.}~\bibnamefont
  {Vahala}}, \ and\ \bibinfo {author} {\bibfnamefont {T.}~\bibnamefont
  {Udem}},\ }\href {\doibase 10.1126/science.aay3676} {\bibfield  {journal}
  {\bibinfo  {journal} {Science}\ }\textbf {\bibinfo {volume} {369}},\ \bibinfo
  {pages} {eaay3676} (\bibinfo {year} {2020})}\BibitemShut {NoStop}%
\bibitem [{\citenamefont {Yao}\ \emph {et~al.}(2016)\citenamefont {Yao},
  \citenamefont {Jiang}, \citenamefont {Wu}, \citenamefont {Yu}, \citenamefont
  {Bi},\ and\ \citenamefont {Ma}}]{M1}%
  \BibitemOpen
  \bibfield  {author} {\bibinfo {author} {\bibfnamefont {Y.}~\bibnamefont
  {Yao}}, \bibinfo {author} {\bibfnamefont {Y.}~\bibnamefont {Jiang}}, \bibinfo
  {author} {\bibfnamefont {L.}~\bibnamefont {Wu}}, \bibinfo {author}
  {\bibfnamefont {H.}~\bibnamefont {Yu}}, \bibinfo {author} {\bibfnamefont
  {Z.}~\bibnamefont {Bi}}, \ and\ \bibinfo {author} {\bibfnamefont
  {L.}~\bibnamefont {Ma}},\ }\href {\doibase 10.1063/1.4963690} {\bibfield
  {journal} {\bibinfo  {journal} {Applied Physics Letters}\ }\textbf {\bibinfo
  {volume} {109}},\ \bibinfo {pages} {131102} (\bibinfo {year}
  {2016})}\BibitemShut {NoStop}%
\bibitem [{\citenamefont {Yao}\ \emph {et~al.}(2021)\citenamefont {Yao},
  \citenamefont {Li}, \citenamefont {Yang}, \citenamefont {Chen}, \citenamefont
  {Hao}, \citenamefont {Yu}, \citenamefont {Jiang},\ and\ \citenamefont
  {Ma}}]{M2}%
  \BibitemOpen
  \bibfield  {author} {\bibinfo {author} {\bibfnamefont {Y.}~\bibnamefont
  {Yao}}, \bibinfo {author} {\bibfnamefont {B.}~\bibnamefont {Li}}, \bibinfo
  {author} {\bibfnamefont {G.}~\bibnamefont {Yang}}, \bibinfo {author}
  {\bibfnamefont {X.}~\bibnamefont {Chen}}, \bibinfo {author} {\bibfnamefont
  {Y.}~\bibnamefont {Hao}}, \bibinfo {author} {\bibfnamefont {H.}~\bibnamefont
  {Yu}}, \bibinfo {author} {\bibfnamefont {Y.}~\bibnamefont {Jiang}}, \ and\
  \bibinfo {author} {\bibfnamefont {L.}~\bibnamefont {Ma}},\ }\href {\doibase
  10.1364/PRJ.409534} {\bibfield  {journal} {\bibinfo  {journal} {Photonics
  Research}\ }\textbf {\bibinfo {volume} {9}},\ \bibinfo {pages} {98} (\bibinfo
  {year} {2021})}\BibitemShut {NoStop}%
\bibitem [{\citenamefont {Gatti}\ \emph {et~al.}(2012)\citenamefont {Gatti},
  \citenamefont {Sala}, \citenamefont {Gambetta}, \citenamefont {Coluccelli},
  \citenamefont {Conti}, \citenamefont {Galzerano}, \citenamefont {Laporta},\
  and\ \citenamefont {Marangoni}}]{R12}%
  \BibitemOpen
  \bibfield  {author} {\bibinfo {author} {\bibfnamefont {D.}~\bibnamefont
  {Gatti}}, \bibinfo {author} {\bibfnamefont {T.}~\bibnamefont {Sala}},
  \bibinfo {author} {\bibfnamefont {A.}~\bibnamefont {Gambetta}}, \bibinfo
  {author} {\bibfnamefont {N.}~\bibnamefont {Coluccelli}}, \bibinfo {author}
  {\bibfnamefont {G.~N.}\ \bibnamefont {Conti}}, \bibinfo {author}
  {\bibfnamefont {G.}~\bibnamefont {Galzerano}}, \bibinfo {author}
  {\bibfnamefont {P.}~\bibnamefont {Laporta}}, \ and\ \bibinfo {author}
  {\bibfnamefont {M.}~\bibnamefont {Marangoni}},\ }\href {\doibase
  10.1364/OE.20.024880} {\bibfield  {journal} {\bibinfo  {journal} {Opt.
  Express}\ }\textbf {\bibinfo {volume} {20}},\ \bibinfo {pages} {24880}
  (\bibinfo {year} {2012})}\BibitemShut {NoStop}%
\bibitem [{\citenamefont {Guo}\ \emph {et~al.}(2020)\citenamefont {Guo},
  \citenamefont {Favier}, \citenamefont {Galland}, \citenamefont {Cambier},
  \citenamefont {Álvarez Martínez}, \citenamefont {Lours}, \citenamefont
  {De~Sarlo}, \citenamefont {Andia}, \citenamefont {Le~Targat},\ and\
  \citenamefont {Bize}}]{Z9}%
  \BibitemOpen
  \bibfield  {author} {\bibinfo {author} {\bibfnamefont {C.}~\bibnamefont
  {Guo}}, \bibinfo {author} {\bibfnamefont {M.}~\bibnamefont {Favier}},
  \bibinfo {author} {\bibfnamefont {N.}~\bibnamefont {Galland}}, \bibinfo
  {author} {\bibfnamefont {V.}~\bibnamefont {Cambier}}, \bibinfo {author}
  {\bibfnamefont {H.}~\bibnamefont {Álvarez Martínez}}, \bibinfo {author}
  {\bibfnamefont {M.}~\bibnamefont {Lours}}, \bibinfo {author} {\bibfnamefont
  {L.}~\bibnamefont {De~Sarlo}}, \bibinfo {author} {\bibfnamefont
  {M.}~\bibnamefont {Andia}}, \bibinfo {author} {\bibfnamefont
  {R.}~\bibnamefont {Le~Targat}}, \ and\ \bibinfo {author} {\bibfnamefont
  {S.}~\bibnamefont {Bize}},\ }\href {\doibase 10.1063/1.5140793} {\bibfield
  {journal} {\bibinfo  {journal} {Rev. Sci. Instrum.}\ }\textbf {\bibinfo
  {volume} {91}},\ \bibinfo {pages} {033202} (\bibinfo {year}
  {2020})}\BibitemShut {NoStop}%
\bibitem [{\citenamefont {Yasui}\ \emph {et~al.}(2021)\citenamefont {Yasui},
  \citenamefont {Hiraishi}, \citenamefont {Ishizawa}, \citenamefont {Omi},
  \citenamefont {Kaji}, \citenamefont {Adachi},\ and\ \citenamefont
  {Tawara}}]{R13}%
  \BibitemOpen
  \bibfield  {author} {\bibinfo {author} {\bibfnamefont {S.}~\bibnamefont
  {Yasui}}, \bibinfo {author} {\bibfnamefont {M.}~\bibnamefont {Hiraishi}},
  \bibinfo {author} {\bibfnamefont {A.}~\bibnamefont {Ishizawa}}, \bibinfo
  {author} {\bibfnamefont {H.}~\bibnamefont {Omi}}, \bibinfo {author}
  {\bibfnamefont {R.}~\bibnamefont {Kaji}}, \bibinfo {author} {\bibfnamefont
  {S.}~\bibnamefont {Adachi}}, \ and\ \bibinfo {author} {\bibfnamefont
  {T.}~\bibnamefont {Tawara}},\ }\href {\doibase 10.1364/OE.433002} {\bibfield
  {journal} {\bibinfo  {journal} {Opt. Express}\ }\textbf {\bibinfo {volume}
  {29}},\ \bibinfo {pages} {27137} (\bibinfo {year} {2021})}\BibitemShut
  {NoStop}%
\bibitem [{\citenamefont {Schibli}\ \emph {et~al.}(2005)\citenamefont
  {Schibli}, \citenamefont {Minoshima}, \citenamefont {Hong}, \citenamefont
  {Inaba}, \citenamefont {Bitou}, \citenamefont {Onae},\ and\ \citenamefont
  {Matsumoto}}]{R20}%
  \BibitemOpen
  \bibfield  {author} {\bibinfo {author} {\bibfnamefont {T.~R.}\ \bibnamefont
  {Schibli}}, \bibinfo {author} {\bibfnamefont {K.}~\bibnamefont {Minoshima}},
  \bibinfo {author} {\bibfnamefont {F.~L.}\ \bibnamefont {Hong}}, \bibinfo
  {author} {\bibfnamefont {H.}~\bibnamefont {Inaba}}, \bibinfo {author}
  {\bibfnamefont {Y.}~\bibnamefont {Bitou}}, \bibinfo {author} {\bibfnamefont
  {A.}~\bibnamefont {Onae}}, \ and\ \bibinfo {author} {\bibfnamefont
  {H.}~\bibnamefont {Matsumoto}},\ }\href {\doibase 10.1364/OL.30.002323}
  {\bibfield  {journal} {\bibinfo  {journal} {Opt. Lett.}\ }\textbf {\bibinfo
  {volume} {30}},\ \bibinfo {pages} {2323} (\bibinfo {year}
  {2005})}\BibitemShut {NoStop}%
\bibitem [{\citenamefont {Benkler}\ \emph {et~al.}(2013)\citenamefont
  {Benkler}, \citenamefont {Rohde},\ and\ \citenamefont {Telle}}]{R14}%
  \BibitemOpen
  \bibfield  {author} {\bibinfo {author} {\bibfnamefont {E.}~\bibnamefont
  {Benkler}}, \bibinfo {author} {\bibfnamefont {F.}~\bibnamefont {Rohde}}, \
  and\ \bibinfo {author} {\bibfnamefont {H.~R.}\ \bibnamefont {Telle}},\ }\href
  {\doibase 10.1364/OE.21.005793} {\bibfield  {journal} {\bibinfo  {journal}
  {Opt. Express}\ }\textbf {\bibinfo {volume} {21}},\ \bibinfo {pages} {5793}
  (\bibinfo {year} {2013})}\BibitemShut {NoStop}%
\bibitem [{\citenamefont {Washburn}\ \emph {et~al.}(2004)\citenamefont
  {Washburn}, \citenamefont {Fox}, \citenamefont {Newbury}, \citenamefont
  {Nicholson}, \citenamefont {Feder}, \citenamefont {Westbrook},\ and\
  \citenamefont {Jørgensen}}]{R15}%
  \BibitemOpen
  \bibfield  {author} {\bibinfo {author} {\bibfnamefont {B.~R.}\ \bibnamefont
  {Washburn}}, \bibinfo {author} {\bibfnamefont {R.~W.}\ \bibnamefont {Fox}},
  \bibinfo {author} {\bibfnamefont {N.~R.}\ \bibnamefont {Newbury}}, \bibinfo
  {author} {\bibfnamefont {J.~W.}\ \bibnamefont {Nicholson}}, \bibinfo {author}
  {\bibfnamefont {K.}~\bibnamefont {Feder}}, \bibinfo {author} {\bibfnamefont
  {P.~S.}\ \bibnamefont {Westbrook}}, \ and\ \bibinfo {author} {\bibfnamefont
  {C.~G.}\ \bibnamefont {Jørgensen}},\ }\href {\doibase
  10.1364/OPEX.12.004999} {\bibfield  {journal} {\bibinfo  {journal} {Opt.
  Express}\ }\textbf {\bibinfo {volume} {12}},\ \bibinfo {pages} {4999}
  (\bibinfo {year} {2004})}\BibitemShut {NoStop}%
\bibitem [{\citenamefont {Peltola}\ \emph {et~al.}(2014)\citenamefont
  {Peltola}, \citenamefont {Vainio}, \citenamefont {Fordell}, \citenamefont
  {Hieta}, \citenamefont {Merimaa},\ and\ \citenamefont {Halonen}}]{R16}%
  \BibitemOpen
  \bibfield  {author} {\bibinfo {author} {\bibfnamefont {J.}~\bibnamefont
  {Peltola}}, \bibinfo {author} {\bibfnamefont {M.}~\bibnamefont {Vainio}},
  \bibinfo {author} {\bibfnamefont {T.}~\bibnamefont {Fordell}}, \bibinfo
  {author} {\bibfnamefont {T.}~\bibnamefont {Hieta}}, \bibinfo {author}
  {\bibfnamefont {M.}~\bibnamefont {Merimaa}}, \ and\ \bibinfo {author}
  {\bibfnamefont {L.}~\bibnamefont {Halonen}},\ }\href {\doibase
  10.1364/OE.22.032429} {\bibfield  {journal} {\bibinfo  {journal} {Opt.
  Express}\ }\textbf {\bibinfo {volume} {22}},\ \bibinfo {pages} {32429}
  (\bibinfo {year} {2014})}\BibitemShut {NoStop}%
\bibitem [{\citenamefont {Watanabe}\ \emph {et~al.}(2017)\citenamefont
  {Watanabe}, \citenamefont {Tamura}, \citenamefont {Musha},\ and\
  \citenamefont {Nakagawa}}]{R17}%
  \BibitemOpen
  \bibfield  {author} {\bibinfo {author} {\bibfnamefont {N.}~\bibnamefont
  {Watanabe}}, \bibinfo {author} {\bibfnamefont {H.}~\bibnamefont {Tamura}},
  \bibinfo {author} {\bibfnamefont {M.}~\bibnamefont {Musha}}, \ and\ \bibinfo
  {author} {\bibfnamefont {K.}~\bibnamefont {Nakagawa}},\ }\href {\doibase
  10.7567/jjap.56.112401} {\bibfield  {journal} {\bibinfo  {journal} {Jpn. J.
  Appl. Phys.}\ }\textbf {\bibinfo {volume} {56}},\ \bibinfo {pages} {112401}
  (\bibinfo {year} {2017})}\BibitemShut {NoStop}%
\bibitem [{\citenamefont {Wu}\ \emph {et~al.}(2012)\citenamefont {Wu},
  \citenamefont {Zhang}, \citenamefont {Wei},\ and\ \citenamefont {Li}}]{Z8}%
  \BibitemOpen
  \bibfield  {author} {\bibinfo {author} {\bibfnamefont {X.}~\bibnamefont
  {Wu}}, \bibinfo {author} {\bibfnamefont {J.}~\bibnamefont {Zhang}}, \bibinfo
  {author} {\bibfnamefont {H.}~\bibnamefont {Wei}}, \ and\ \bibinfo {author}
  {\bibfnamefont {Y.}~\bibnamefont {Li}},\ }\href {\doibase 10.1063/1.4737625}
  {\bibfield  {journal} {\bibinfo  {journal} {Rev. Sci. Instrum.}\ }\textbf
  {\bibinfo {volume} {83}},\ \bibinfo {pages} {073107} (\bibinfo {year}
  {2012})}\BibitemShut {NoStop}%
\bibitem [{\citenamefont {Jost}\ \emph {et~al.}(2002)\citenamefont {Jost},
  \citenamefont {Hall},\ and\ \citenamefont {Ye}}]{R18}%
  \BibitemOpen
  \bibfield  {author} {\bibinfo {author} {\bibfnamefont {J.~D.}\ \bibnamefont
  {Jost}}, \bibinfo {author} {\bibfnamefont {J.~L.}\ \bibnamefont {Hall}}, \
  and\ \bibinfo {author} {\bibfnamefont {J.}~\bibnamefont {Ye}},\ }\href
  {\doibase 10.1364/OE.10.000515} {\bibfield  {journal} {\bibinfo  {journal}
  {Opt. Express}\ }\textbf {\bibinfo {volume} {10}},\ \bibinfo {pages} {515}
  (\bibinfo {year} {2002})}\BibitemShut {NoStop}%
\bibitem [{\citenamefont {Fordell}\ \emph {et~al.}(2014)\citenamefont
  {Fordell}, \citenamefont {Wallin}, \citenamefont {Lindvall}, \citenamefont
  {Vainio},\ and\ \citenamefont {Merimaa}}]{R19}%
  \BibitemOpen
  \bibfield  {author} {\bibinfo {author} {\bibfnamefont {T.}~\bibnamefont
  {Fordell}}, \bibinfo {author} {\bibfnamefont {A.~E.}\ \bibnamefont {Wallin}},
  \bibinfo {author} {\bibfnamefont {T.}~\bibnamefont {Lindvall}}, \bibinfo
  {author} {\bibfnamefont {M.}~\bibnamefont {Vainio}}, \ and\ \bibinfo {author}
  {\bibfnamefont {M.}~\bibnamefont {Merimaa}},\ }\href {\doibase
  10.1364/AO.53.007476} {\bibfield  {journal} {\bibinfo  {journal} {Appl.
  Opt.}\ }\textbf {\bibinfo {volume} {53}},\ \bibinfo {pages} {7476} (\bibinfo
  {year} {2014})}\BibitemShut {NoStop}%
\bibitem [{\citenamefont {Gunton}\ \emph {et~al.}(2015)\citenamefont {Gunton},
  \citenamefont {Semczuk},\ and\ \citenamefont {Madison}}]{R21}%
  \BibitemOpen
  \bibfield  {author} {\bibinfo {author} {\bibfnamefont {W.}~\bibnamefont
  {Gunton}}, \bibinfo {author} {\bibfnamefont {M.}~\bibnamefont {Semczuk}}, \
  and\ \bibinfo {author} {\bibfnamefont {K.~W.}\ \bibnamefont {Madison}},\
  }\href {\doibase 10.1364/OL.40.004372} {\bibfield  {journal} {\bibinfo
  {journal} {Opt. Lett.}\ }\textbf {\bibinfo {volume} {40}},\ \bibinfo {pages}
  {4372} (\bibinfo {year} {2015})}\BibitemShut {NoStop}%
\bibitem [{\citenamefont {Prehn}\ \emph {et~al.}(2017)\citenamefont {Prehn},
  \citenamefont {Glockner}, \citenamefont {Rempe},\ and\ \citenamefont
  {Zeppenfeld}}]{R22}%
  \BibitemOpen
  \bibfield  {author} {\bibinfo {author} {\bibfnamefont {A.}~\bibnamefont
  {Prehn}}, \bibinfo {author} {\bibfnamefont {R.}~\bibnamefont {Glockner}},
  \bibinfo {author} {\bibfnamefont {G.}~\bibnamefont {Rempe}}, \ and\ \bibinfo
  {author} {\bibfnamefont {M.}~\bibnamefont {Zeppenfeld}},\ }\href {\doibase
  10.1063/1.4977049} {\bibfield  {journal} {\bibinfo  {journal} {Rev. Sci.
  Instrum.}\ }\textbf {\bibinfo {volume} {88}},\ \bibinfo {pages} {033101}
  (\bibinfo {year} {2017})}\BibitemShut {NoStop}%
\bibitem [{\citenamefont {Donley}\ \emph {et~al.}(2005)\citenamefont {Donley},
  \citenamefont {Heavner}, \citenamefont {Levi}, \citenamefont {Tataw},\ and\
  \citenamefont {Jefferts}}]{R23}%
  \BibitemOpen
  \bibfield  {author} {\bibinfo {author} {\bibfnamefont {E.~A.}\ \bibnamefont
  {Donley}}, \bibinfo {author} {\bibfnamefont {T.~P.}\ \bibnamefont {Heavner}},
  \bibinfo {author} {\bibfnamefont {F.}~\bibnamefont {Levi}}, \bibinfo {author}
  {\bibfnamefont {M.~O.}\ \bibnamefont {Tataw}}, \ and\ \bibinfo {author}
  {\bibfnamefont {S.~R.}\ \bibnamefont {Jefferts}},\ }\href {\doibase
  10.1063/1.1930095} {\bibfield  {journal} {\bibinfo  {journal} {Review of
  Scientific Instruments}\ }\textbf {\bibinfo {volume} {76}},\ \bibinfo {pages}
  {063112} (\bibinfo {year} {2005})}\BibitemShut {NoStop}%
\end{thebibliography}%
 
 \end{document}